\documentclass[english,prl,twocolumn,superscriptaddress,amsmath,amssymb,a4paper]{revtex4}
\usepackage[T1]{fontenc}
\usepackage[latin1]{inputenc}
\usepackage{graphicx}
\usepackage[top=0.85in,bottom=0.9in,,left=0.75in,right=0.75in]{geometry}
\usepackage{booktabs}
\usepackage{color}
\usepackage{verbatim}

\makeatletter
\usepackage{dcolumn}
\usepackage{bm}
\usepackage{epsfig}

\newcommand{\ignore}[1]{}

\usepackage{babel}
\makeatother

\usepackage{sidecap}
\sidecaptionvpos{figure}{c}

\begin{document}

\title{Enhanced Itinerant Ferromagnetism in Hole-doped Transition Metal Oxides: Beyond the Canonical Double Exchange Mechanism}

\author{Zhao Liu}
\email{Zhao.Liu@monash.edu}
\affiliation{Department of Materials Science and Engineering, Monash University, Victoria 3800, Australia}
\affiliation{ARC Centre of Excellent in Future Low-Energy Electronics Technologies, Monash University, Victoria 3800, Australia}

\author{Nikhil V. Medhekar}
\email{Nikhil.Medhekar@monash.edu}
\affiliation{Department of Materials Science and Engineering, Monash University, Victoria 3800, Australia}
\affiliation{ARC Centre of Excellent in Future Low-Energy Electronics Technologies, Monash University, Victoria 3800, Australia}

%\date{\today}% It is always \today, today, % but any date may be explicitly specified

\begin{abstract}
Here we demonstrate the occurrence of robust itinerant ferromagnetism in Mott-Hubbard systems at both low and high doping concentrations. Specifically, we study the effect of hole doping on the experimentally synthesized LaCrAsO via first-principles calculations and observe that the parent G-type antiferromagnetism vanishes quickly at low doping concentration ($x$ $\sim$ 0.20) and the system becomes ferromagnetic metal due to the canonical double exchange (CDE) mechanism. As $x$ continues to increase, the onsite energy difference between Cr 3$d$ and As 4$p$ orbitals decreases and the system transitions to a ferromagnetic negative charge-transfer energy metal. Therefore, the itinerant ferromagnetism doesn't terminate at intermediate $x$ as CDE mechanism usually predicts. Furthermore, our calculations reveal that both nearest and next-nearest ferromagnetic exchange coupling strengths keep growing with $x$, showing that ferromagnetism caused by negative charge-transfer energy state is "stronger" than that of CDE picture. Our work not only unveils an alternative mechanism of itinerant ferromagnetism, but also has the potential to attract immediate interest among experimentalists.  
\end{abstract}

\maketitle

\textit{Introduction.}-One of the oldest but recurrent topics in condensed matter physics is itinerant ferromagnetism (FM) \cite{Moriya1985, Belitz2005, Katsnelson2008, Brando2016, Li2014, Deng2018, Bonilla2018, Sharpe2019, Park2021, Bao2022}. Discovered since ancient time in materials such as elemental iron and nickel, itinerant FM now plays a crucial role in various technological applications, including modern-day data processing and storage \cite{Prinz1998}. Several mechanisms have been proposed for itinerant FM, including Stoner criterion \cite{Stoner1938}, Nagaoka's theorem \cite{Nagaoka1966, Tasaki1989, Bobrow2018}, flat-band model \cite{Mielke1991_1, Mielke1991_2, Tasaki1992, Mielke1993}, multi-orbital Hubbard models \cite{Shen1989, Sakai2007, LiY2014}, canonical double exchange (CDE) \cite{Zener1951, Anderson1955}, etc. Among these mechanisms, CDE is a prominent one in both mixed valence transition metal oxides (such as magnetites \cite{Hu2002} and spinel ferrites \cite{Ramirez1997}) and doped Mott-Hubbard insulators \cite{Khomskii2014}. In the latter system, itinerant charge carriers arise from cation's broad $d$ bands while the anion's $p$ orbitals remain inactive. To maximize the kinetic energy, an antiferromagnetism (AFM)-FM transition occurs at low doping concentration. In manganite perovskites, the occurrence of itinerant FM is marked by a strong suppression of resistivity by magnetic field during the FM-paramagnetism phase transition \cite{Cheong1999}. This is now known as the colossal negative magnetoresistance, a crucial concept in spintronics. Nevertheless, a transition from FM to AFM always occurs at high doping concentration, limiting the application of itinerant FM in spintronics.

This work introduces a mechanism for persistent itinerant FM that can survive at high doping concentration in Mott-Hubbard systems. To set the stage, we first discuss the prototypical Mott-Hubbard system: LaCrAsO, which has been experimentally synthesized \cite{Park2013}. While LaCrAsO is metallic, several indirect observations suggest that it is in close proximity to a Mott-Hubbard insulator \cite{Park2013}. Firstly, the measured room-temperature electrical resistivity $\rho$ $\sim$ 3.8 m$\Omega \cdot cm$, which corresponds to a normalized mean free path $k_Fl \sim hc/e^2\rho \sim $ 0.6 (where c = 8.98 {\AA} is the lattice constant along the c direction). The fact that $k_F l$ < 1 strongly indicates LaCrAsO is a bad metal. Secondly, a local magnetic momentum of 1.57 $\mu_B$/Cr is reported at room temperature---the existence of local magnetic momenta implies a strong correlation effect. Finally, the ground state of LaCrAsO is G-type AFM, such a long-range order is the result of superexchange mechanism. Taking all of these experimental results together, even if LaCrAsO is not fully Mott-Hubbard insulating, it should not be far away from it. Consequently, various theoretical studies have explored the possibility of high-temperature superconductivity in electron doped LaCrAsO \cite{Pizarro2017, Wang2017} and BaCr$_2$As$_2$ \cite{Edelmann2017}, aiming at constructing a similar 3$d$ orbital filling to Fe$^{2+}$ in the well-known superconductor LaFeAsO \cite{Kamihara2008}. 

Here we investigate the other type of charge doping: hole doping (its concentration is labelled by $x$), where the driving force is to upgrade the valence state of Cr from 2+ to 3+. As described earlier, the CDE mechanism can induce an AFM-FM transition, but only at small $x$, and at large $x$ the system often returns to AFM phase as the superexchange mechanism dominates. On the other hand, it is known that the higher valence state Cr$^{3+}$ can lead to negative charge-transfer energy states which greatly enhance FM \cite{Khomskii1997, Liu2023}. Based on these arguments, there will be a competition between superexchange mechanism and negative charge-transfer energy states in heavily hole-doped LaCrAsO, which makes it intriguing to explore the evolution of magnetic ground state as well as the exchange coupling strength ($J$) with respect to $x$. 

%\begin{SCfigure*}
\begin{figure*}
\includegraphics[width=18cm]{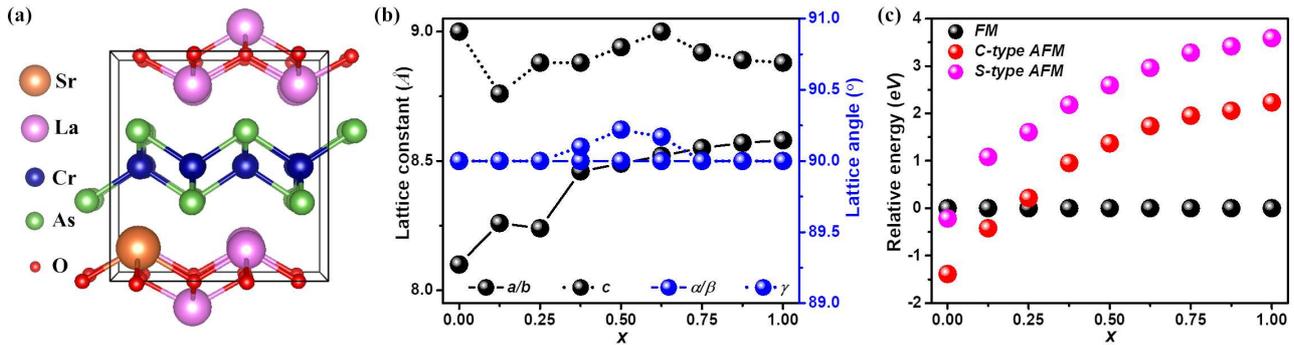}
\caption{(a) Perspective view of 2 $\times$ 2 $\times$ 1 (La$_{1-x}$Sr$_x$)CrAsO supercell, here $x$ = 0.125 is shown as an example. (b) Lattice constants a, b, c (black color) and angles $\alpha$, $\beta$, $\gamma$ (blue color) of the ground state structures at different $x$ values. (c) Relative energies of three long-range magnetic orders at different $x$ values.}
\label{FIG-1}
\end{figure*}
%\end{SCfigure*}

In this work, via first-principles calculations, we demonstrate that the parent G-type AFM quickly disappears at small $x$ values, as predicted by the CDE mechanism. However, contrary to CDE mechanism, the FM doesn't vanish at an intermediate $x$ value but persists up to $x$ = 1.00, violating the superexchange mechanism. Additionally, our calculations reveal that the nearest and next-nearest ferromagnetic exchange coupling strengths (labelled as $J_1$ and $J_2$) continue to grow with increasing $x$, indicating a stronger itinerant FM than that of the CDE mechanism. We attribute this enhanced itinerant FM at large $x$ values to the formation of negative charge-transfer energy states . 

\textit{First-principles calculations.}-The first-principles calculations were performed using the Vienna \textit{ab-initio} simulation package (VASP) within the framework of density functional theory (DFT) \cite{VASP}. For geometric optimization and electronic property calculations, a plane-wave cutoff 600 eV was used. All our calculations were converged within 10$^{-5}$ eV for energy and 0.01 eV/{\AA} for Hellman-Feynman forces. For the LaCrAsO unitcell, the Brillouin zone integration was carried out with 14 $\times$ 14 $\times$ 10 k-point sampling for self-consistency. The majority of calculations were based on the non-empirical, strongly constrained and appropriately normed (SCAN) functional \cite{Sun2015, Sun2016}, which is a parameter-free functional and can treat charge, spin and lattice degrees of freedom on equal footing. For the interlayer van der Waals (vdW) interactions, the revised Vydrov-van Voorhis nonlocal correlation functional ($r$VV10) was employed \cite{Peng2016}. It is found that SCAN + $r$VV10 shows a good performance on LaCrAsO by reproducing several key experimental results (see Sec. A of Supplementary Materials \cite{SI} for details). In addition, we also employed the hybrid HSE06 functional \cite{HSE2003, HSE2004} to correct the magnetic band structure.  

The hole doping in LaCrAsO is achieved by partially replacing La with Sr in a supercell. The ionic size of Sr$^{2+}$ is almost identical to that of La$^{2+}$, thus, a true solid solution should be formed over a large $x$ range. This substitution method has been widely used in hole-doped materials, such as cuprates \cite{Bednorz1986}, iron-based pnictides \cite{Wen2008}, infinite-layer nickelate \cite{Li2019}, and  magnanites \cite{Cheong1999}. Since charge/spin/orbital anomalies have been reported at doping concentration $x$ = n/8 (n is an integer) for both square lattice based cuprates \cite{Moodenbaugh1988, Tranquada1995} and perovskite based magnanites \cite{Cheong1999}, a 2 $\times$ 2 $\times$ 1 supercell was adopted in this work (see {\color{blue} Fig. \ref{FIG-1}(a)}). By replacing n (n = 1--7) La atoms with Sr, $x$ value from 0.00 to 1.00 can be simulated. To account for any possible distortion introduced by hole doping, both the supercell and atomic coordinates were allowed to optimize freely. For all $x$ values, our calculations show that (La$_{1-x}$Sr$_x$)CrAsO alloy is energetically preferred (see Sec. B of Supplementary Materials \cite{SI}). The optimized lattice constants and angles of the ground states are presented in {\color{blue} Fig. \ref{FIG-1}(b)}, from which it is clear that all the supercells have negligible distortions as the three angles $\alpha$, $\beta$ and $\gamma$ are all close to 90$^{\circ}$. Such small distortions can be traced back to the similar ionic radius of La$^{3+}$ and Sr$^{2+}$.  To provide a better description of the electronic structures at $x$ = 0.50, it was also simulated in a unitcell.

\textit{Results and discussion.}-First we explore the effect of hole doping on the fundamental electronic structures and in particular, the charge-transfer energy. The charge-transfer energy composes of two parts: single-particle part $\varepsilon_{dp}$ which is the onsite energy difference between Cr 3$d$ and As 4$p$ orbitals, and interacting part E$_{int}$ which is a function of the interaction parameters and electron fillings. As shown in {\color{blue} Fig. S3}, as $x$ increases, the position of As 4$p$ orbitals gradually shift to higher energy compared with Cr 3$d$, leading to a reduced $\varepsilon_{dp}$. Such a phenomenon has also been reported in hole doped infinite-layer nickelate recently \cite{Liu2021}. Additionally, the charge-transfer energy of an electron from As 4$p$ to Cr 3$d$ orbitals includes the Hubbard repulsion of the transferred electron with the $d$-electrons already present on the Cr ions. Therefore, E$_{int}$ is reduced with a decrease in the number of $d$-electrons, following approximately E$_{int}$(Cr$^{3+}$) = E$_{int}$(Cr$^{2+}$) - U$_d$, where U$_d$ is the average Hubbard interaction of Cr 3$d$ orbitals \cite{Khomskii2014}. Taken together, with high cationic valence states, charge-transfer energy is significantly reduced. 

%\begin{SCfigure*}
\begin{figure*}
\includegraphics[width=16cm]{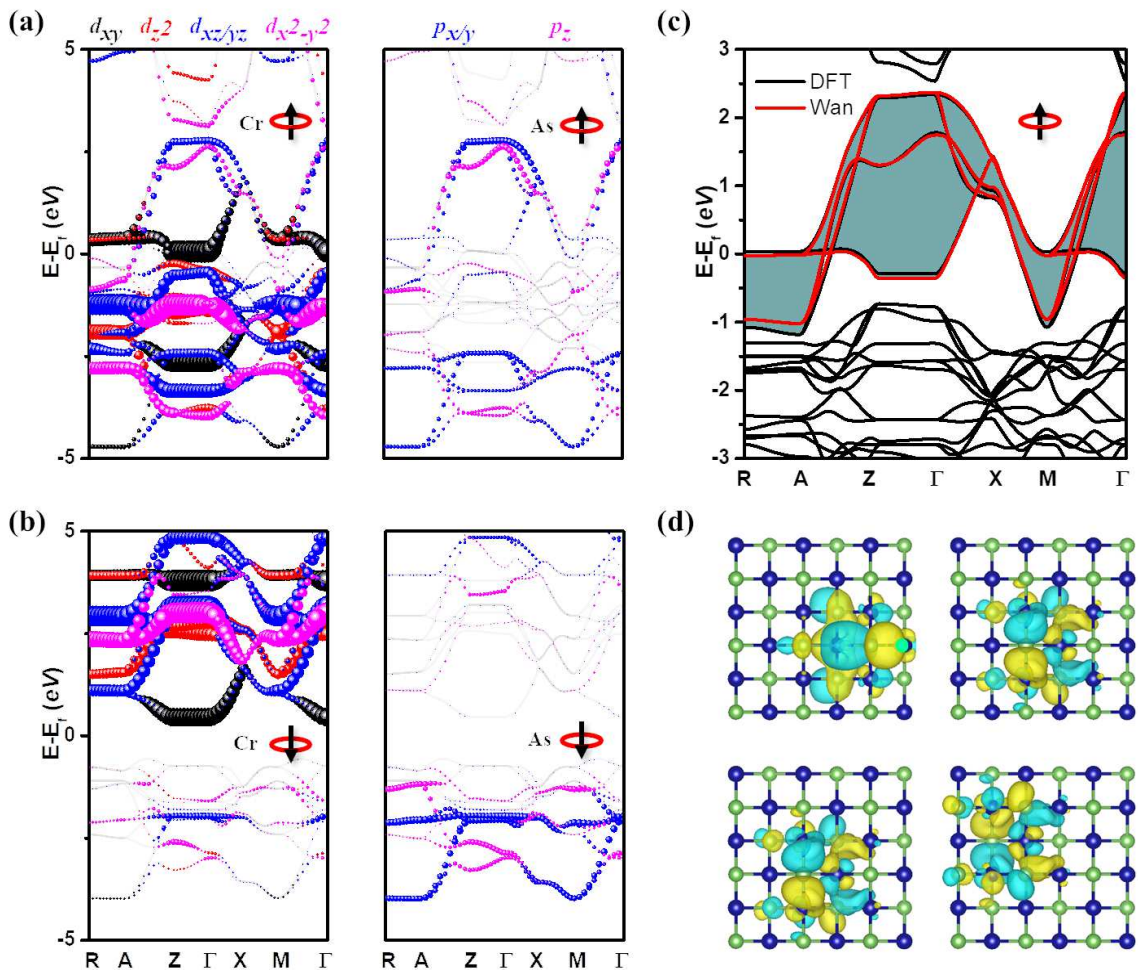}
\caption{
(a,b) Orbital-resolved magnetic band structure of SrCrAsO for spin up and spin down channel. The high symmetry \textbf{k}-path is R-A-Z-$\Gamma$-X-M-$\Gamma$: (0.5, 0.5, 0.0)-(0.5, 0.5, 0.5)-(0.0, 0.0, 0.5)-(0.0, 0.0, 0.0)-(0.5, 0.0, 0.0)-(0.5, 0.5, 0.0)-(0.0, 0.0, 0.0). (c) Magnetic band structure of (La$_{0.5}$Sr$_{0.5}$)CrAsO (in a unitcell) in the spin up channel (black). Red: the Wannier fitted band structure. (d) Topview of the four maximally localized Wannier functions downfolded from the grey color shaded bands in (c). The (La$_{0.5}$Sr$_{0.5}$)O sublayer is omitted for a better view and the isovalue is $\pm$ 0.60 {\AA}$^{-3/2}$.
}
\label{FIG-2}
\end{figure*}
%\end{SCfigure*}

To investigate the magnetic ground state, three long-range magnetic orders were considered: FM with magnetic ordering momentum $\textbf{q}$ = (0, 0, 0), checkboard AFM (C-type AFM) with $\textbf{q}$ = ($\pi$, $\pi$, 0) and strip AFM (S-type AFM) with $\textbf{q}$ = ($\pi$, 0, 0). The relative energies of these magnetic orders at different $x$ values are shown in {\color{blue} Fig. \ref{FIG-1}(c)}. At $x$ = 0.00, C-type AFM is the ground state, which is ascribed to the superexchange mechanism (see Sec. D of Supplementary Materials \cite{SI} for more details). This magnetic order persists up to $x \sim$ 0.20, beyond which the FM order becomes the lowest one. As $x$ further increases, the energy difference between FM and other AFM magnetic order grows larger, in accordance with the CDE picture. However, it is unexpected that the energy difference continues to grow without saturating even at $x$ = 1.00. This behavior significantly deviates from the CDE picture in the sense that there is no optimal doping concentration ($x_c$). In the CDE mechanism, the gain from kinetic energy reaches the most at $x_c$ with $T_c$ the highest. After $x_c$, the gain from exchange energy gradually prevails and the superexchange mechanism should make SrCrAsO AFM (see Sec. D of Supplementary Materials \cite{SI} for details). In order to explain this unconventional doping behaviour, we now turn to the electronic structures. 

%\begin{SCfigure*}
\begin{figure*}
\includegraphics[width=17cm]{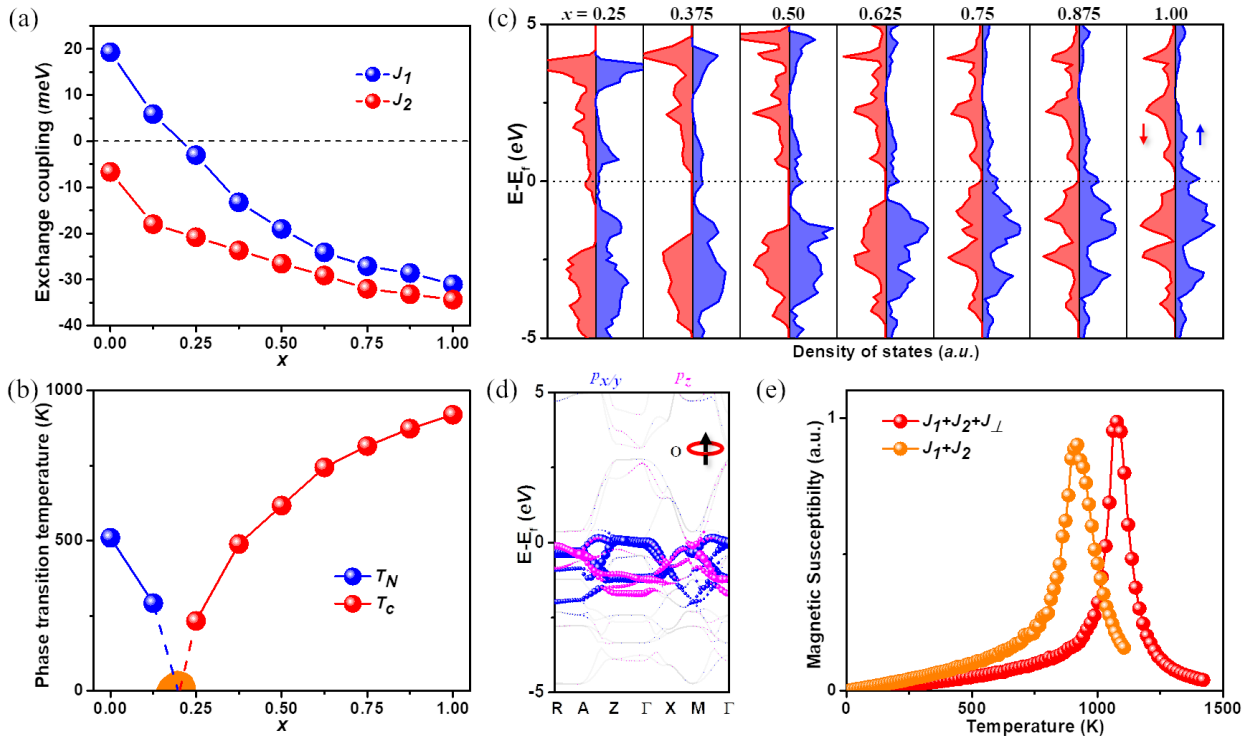}
\caption{(a) $J_1$ and $J_2$ at different $x$ values. (b) Phase transition temperature at different $x$ values. The orange colored region near $x$ = 0.20 is the critical point region. (c) Total density of states at $x$ > 0.20, red/blue represents spin dn/up channel respectively. (d) O 2$p$ orbital-resolved magnetic band structure of SrCrAsO in spin up channel. (e) Evolution of magnetic susceptibility with respect to temperature for different Heisenberg model in SrCrAsO.}
\label{FIG-3}
\end{figure*}
%\end{SCfigure*}  

{\color{blue} Fig. \ref{FIG-2}(a)-(b)} presents the orbital- and spin-resolved FM band structures of SrCrAsO ($x$ = 1.00). We observe that the spin up channel is metallic while spin down channel is insulating, indicating that SrCrAsO is a half metal (HSE06 functional also confirms this, see {\color{blue} Fig. S7}). Moreover, there is a significant disparity in orbital compositions for the two spin channels. Specifically, in the spin down channel, the valence and conduction bands around Fermi level (E$_f$) are contributed by As 4$p$/O 2$p$ and Cr 3$d$ orbitals respectively. Nevertheless, in the spin up channel, there are states of predominantly As 4$p$ character above E$_f$ along Z-$\rm \Gamma$-X direction at around 2.60 eV. We stress that these As 4$p$ states are not originated from $d-p$ hybridization. In the case here, the 4$p$ states would be locating at E$_f$ even with the hybridization switching off, indicating the negative charge-transfer energy nature \cite{Liu2023}. These negative charge-transfer energy states make the magnetic molecular orbitals as the underlying building blocks rather than the localized atomic orbitals, resulting in a large spreading of magnetic orbitals. This, in turn, strongly enhances the FM exchange interaction, as both $J_1$ and $J_2$ will become FM \cite{Liu2023}.

To visualize the magnetic molecular orbitals, we focus on the system at $x$ = 0.50 in an unitcell instead of $x$ = 1.00 for there is no local gap structure around E$_f$ in Fig. \ref{FIG-2}(a). In {\color{blue} Fig. \ref{FIG-2}(c)}, we observe four bands (grey color shaded) are isolated from the others, which can be downfolded to obtain maximally localized Wannier orbitals (MLWFs) \cite{Wanpac}. The overall downfolding is satisfactory as evidenced by the good agreement between the DFT and the Wannier fitted bands (see {\color{blue} Fig. \ref{FIG-2}(c)}). {\color{blue} Fig. \ref{FIG-2}(d)} displays the four MLWFs, and it is clear that they are composed of small CrAs clusters rather than localized atomic orbitals. This feature can be reflected in the magnetic form factor through inelastic neutron scattering, as observed in itinerant chiral magnet MnSi recently \cite{Jin2023}. Similar to the FM CrAs monolayer, these MLWFs are of anti-bonding type, so in principle, with further hole doping, the FM phase should become stronger as the occupation of anti-bonding orbitals decreases \cite{Liu2023}. This actually explains the anomaly we observe in {\color{blue} Fig. \ref{FIG-1}(c)}, where the FM order becomes increasingly stable as $x$ approaches 1.00. Hence, at large $x$, (La$_{1-x}$Sr$_x$)CrAsO becomes a FM negative charge-transfer energy metal and it is noted that the itinerant FM here can't be explained by other exchange mechanisms (see Sec. D of  Supplementary Materials \cite{SI} for detailed elaboration).

After understanding the origin of FM in both small and large $x$ values, we next evaluate the phase transition temperatures ($T_N/T_c$ for Neel/Curie temperature respectively). By mapping the relative energies of FM, C-type AFM and S-type AFM to the Heisenberg model with $S$ = 3/2 (see \cite{SI} for detailed information), both $J_1$ and $J_2$ can be obtained as shown in {\color{blue} Fig. \ref{FIG-3}(a)}. At $x$ = 0.00 and 0.125, $J_1$ is AFM (positive value) while $J_2$ is FM (negative value), therefore C-type AFM is the ground state as confirmed by the DFT calculations. At $x$ $\sim$ 0.20, $J_1$ changes its sign and becomes FM, and then both $J_1$ and $J_2$ are FM. As $x$ increases, the magnitude of both $J_1$ and $J_2$ increase without saturation. As mentioned before, such a feature stems from the fact that the system is closer to "ideal filling" as $x$ approaches 1.00. To determine $T_N/T_c$, classical Monte Carlo (MC) simulations were preformed on a 40 $\times$ 40 $\times$ 1 supercell based on Heisenberg Hamiltonian with $J_1$ and $J_2$ \cite{PASP}. The phase transition temperatures obtained are presented in {\color{blue} Fig. \ref{FIG-3}(b)}. Initially, $T_N$ gradually decreases until a critical point region (labelled by orange color). After that $T_c$ continuously increases to $\sim$ 920 K at $x$ = 1.00. The total density of states for $x$ > 0.20 are displayed in {\color{blue} Fig. \ref{FIG-3}(c)}.  It is evident that most of these FM phases are half metal, except for $x$ = 0.25. The half-metallic gap reaches its maximum value of $\sim$ 1.88 eV at $x$ = 0.375 and reduces to $\sim$ 0.80 eV at $x$ = 1.00. If HSE06 functional is further considered, there is a $\sim$ +0.80 eV correction to the half-metallic gap (see {\color{blue} Fig. S7}). The high $T_c$ and the large half-metallic gap makes (La$_{1-x}$Sr$_x$)CrAsO ($x$ > 0.25) a promising candidate for half-metallic ferromagnets. 

In the above discussions, we concentrated solely on the intralayer coupling. However, we found that at high $x$ values, not only As 4$p$, but also O 2$p$ orbitals are polarized. In {\color{blue} Fig. \ref{FIG-3}(d)}, we present the O 2$p$ orbital-resolved magnetic band structure in the spin up channel, and it is obvious that O 2$p$ bands cross E$_f$. This means that these O 2$p$ states can be regarded as electron or hole reservoirs between CrAs sublayers and promote a three-dimensional (3D) magnet behavior. To investigate the 2D-3D crossover at $x$ = 1.00, we calculated the nearest out-of-plane exchange coupling ($J_{\perp}$) to be -6.40 meV and simulated the magnetic susceptibility with/without $J_{\perp}$, as plotted in {\color{blue} Fig. \ref{FIG-3}(e)}. The introduction of $J_{\perp}$ not only pushes $T_c$ up to a higher temperature ($\sim$ 1080 K), but also gives a sharper peak with smaller full width at half maximum, suggesting the 3D nature of SrCrAsO. 

Because As 4$p$ orbitals play a vital role in stabilizing the FM order, lastly we discuss how to track them experimentally. The contribution of As 4$p$ orbitals can be identified in both energy space and real space. In energy space, a "shoulder" or even a "peak" structure will occur at As $L_{2/3}$ edge in the electron energy loss spectroscopy as $x$ increases, just like the case in cuprates \cite{Romberg1990}. In real space, the valence charge density around the As site should gradually reduce with increasing $x$, which can be observed by synchrotron X-ray diffraction. We also note that similar observations have been made for the ligand hole in cubic perovskite SrFeO$_3$ in a recent study \cite{Kitou2023}. 

\textit{Conclusions.}-
In summary, we propose an alternative phase diagram for doped Mott-Hubbard system. As the concentration of hole doping increases, the valence state of the metallic ions becomes higher and the charge-transfer gap gradually turns to negative. The negative charge-transfer energy states result in strong ferromagnetism, causing the overdoped system to be ferromagnetic rather than antiferromagnetic described by superexchange mechanism. Furthermore, the ferromagnetism at large doping concentration is so robust that it is even stronger than that of canonical double exchange mechanism at low doping. Based on this proposed phase diagram, we expect that a colossal negative magnetoresistance will occur at large doping concentration, similar to that observed in double exchange systems. Therefore, our work not only facilitates the discovery of half-metallic ferromagnets but also expands the search for materials with colossal negative magnetoresistance, which has important practical applications.

We thank J.-W. Li, G. Su and X. Cui for useful discussions. Z. L. and N. V. M. gratefully acknowledge the support from National Computing Infrastructure, Pawsey Supercomputing Facility and the Australian Research Council's Centre of Excellence in Future Low-Energy Electronic Technologies (CE170100039).

\end{document}